\documentclass[prd,twocolumn,amsmath,showpacs,amssymb,superscriptaddress, floatfix, nofootinbib,10pt]{revtex4}
\usepackage{indentfirst}
\usepackage{tikz}
\usepackage{setspace}
\usepackage{amsfonts,color,xcolor,graphicx,amsfonts,multirow,amsmath}
\usepackage[section]{placeins}
\definecolor{bluee}{rgb}{0,0,1}
\usepackage[colorlinks=true, colorlinks=true, citecolor=bluee, linkcolor=bluee, urlcolor=bluee]{hyperref}

\definecolor{darkerblue}{rgb}{0.1, 0.1, 0.7}
\definecolor{darkergreen}{rgb}{0.1, 0.5, 0.1}
\definecolor{lightred}{rgb}{1, 0, 0}
\usepackage{ulem}

\allowdisplaybreaks

\usepackage{slashed}

\begin{document}
\title{Prospective analysis of CKM element $|V_{cd}|$ and $D^+$-meson decay constant from leptonic decays $D^+ \to \ell^+ \nu$}
\author{Ya-Xiong Wang}
\author{Hai-Jiang Tian}
\author{Yin-Long Yang}
\affiliation{Department of Physics, Guizhou Minzu University, Guiyang 550025, P.R.China}
\author{Tao Zhong}
\author{Hai-Bing Fu}
\email{fuhb@gzmu.edu.cn}
\affiliation{Department of Physics, Guizhou Minzu University, Guiyang 550025, P.R.China}
\affiliation{Institute of High Energy Physics, Chinese Academy of Sciences, Beijing 100049, P.R.China}

\begin{abstract}
The leptonic decay of $D^+$-meson has attracted significant interest due to its unique characteristics. In this paper, we carry out an investigation into the $D^+$-meson leptonic decays $D^+\to \ell^+\nu_{\ell}$ with $\ell=(e,\mu,\tau)$ by employing the QCD sum rules approach. In which the $D^+$-meson decay constant $f_{D^+}$ is an important input parameter in the process. To enhance the accuracy of our calculations for $f_{D^+}$, we consider the quark propagator and vertex up to dimension-six within the framework of background field theory. Consequently, we obtain the QCD sum rule expression for $f_{D^+}$ up to dimension-six condensates, yielding $f_{D^+}=203.0\pm1.5~\mathrm{MeV}$. Our results are in good agreement with BESIII measurements and theoretical predictions. We also present the integrated decay widths for the $D^+$-meson in three channels $\Gamma(D^+\to e^+\nu_e)=(5.263_{-0.075}^{+0.076})\times10^{-21}~\mathrm{GeV}$, $\Gamma(D^+\to \mu^+\nu_{\mu})=(2.236_{-0.032}^{+0.032})\times10^{-16}~\mathrm{GeV}$ and $\Gamma(D^+\to \tau^+\nu_{\tau})=(5.958_{-0.085}^{+0.086})\times10^{-16}~\mathrm{GeV}$. Accordingly, we compute the branching fraction $\mathcal{B}(D^+\to\ell^+\nu_{\ell})$ with the electron, muon and tau channels, which are $\mathcal{B}(D^+\to e^+\nu_e)=(8.260_{-0.118}^{+0.119})\times10^{-9}$, $\mathcal{B}(D^+\to\mu^+\nu_{\mu})=(3.508_{-0.050}^{+0.051})\times10^{-4}$ and $\mathcal{B}(D^+\to\tau^+\nu_{\tau})=(0.935_{-0.013}^{+0.013})\times10^{-3}$. Furthermore, we present our prediction for the CKM matrix element $|V_{cd}|$ using the branching fraction $\mathcal{B}(D^+\to\mu^+\nu_{\mu})$ obtained from BESIII Collaboration, yielding $|V_{cd}|=0.227_{-0.001}^{+0.002}$.
\end{abstract}
\date{\today}

\maketitle

\textit{Introduction.--}In the Standard Model (SM), the Cabibbo-Kobayashi-Maskawa (CKM) matrix elements are fundamental parameters, therefore their precision measurements is important and essential~\cite{Kobayashi:1973fv,Cabibbo:1963yz}. Meanwhile, charged mesons, which are formed from a quark and an antiquark, can decay into a lepton pair when these particles annihilate via a $W$-boson. For $D^+_{(s)}$ and $B^+$-mesons, the processes of the quark-antiquark annihilation can be realized via the decay of a virtual $W^+$-boson to the $\ell^+ \nu$ final states.
Measurements of the purely leptonic branching fraction and lifetimes allow an experimental determination of the product of the CKM matrix and the decay constant. The calculations of the charm-light and charm-strange decay constants $f_{D}$ and $f_{D_s}$, respectively, together with the experimentally measured decay rates of the weak decays of $D$ and $D_s$-mesons into a lepton and a neutrino can be used to determine the CKM matrix elements $|V_{cd}|$ and $|V_{cs}|$. However, the structure of the SM requires the CKM matrix elements to be unitary. Thus, through independent and precise determinations of the CKM matrix elements, their orthogonality can be tested, and tests of the SM can be carried out~\cite{Boyle:2018knm}.

In the case of the $B$-mesons, their leptonic decays are among the simplest flavour-changing processes and a very clean processes since all of the nonperturbative, hadronic physics are contained within the decay constant $f_B$~\cite{Black:2022eph, Colquhoun:2015oha}. For the $B$-meson, the experimental studies have yielded a lot of results compared to the $D$-meson in recent years~\cite{Belle-II:2024ami,LHCb:2023ssl,Belle:2023asx,Belle-II:2023qyd}. Therefore, it is meaningful to carry out a study on the leptonic decay of $D^+$-meson to provide a certain reference for future experimental detection. Similarly, the leptonic decay $D^+ \to \ell^+ \nu_{\ell}$ with $\ell = (e, \mu, \tau)$ of $D^+$-mesons is also one of the vital processes in SM, and it plays a significant role in the extraction of the CKM matrix elements $|V_{cd}|$. For the $c\to d$ flavour-changing quark transition, leptonic decay of $D^+$-meson is a good opportunity to examine both strong and weak effects in the charm sector. Accurate non-perturbative calculations of the decay constants combined with precise experimental results provide excellent opportunities to perform tests of lepton flavour universality.

The combination of precise experimental measurements and theoretical predictions allows to extract CKM matrix elements or constrain flavor changing processes in the SM. In 2009, the CLEO Collaboration used the entire CLEO-c $\psi(3770)\to D\bar{D}$ event sample, corresponding to an integrated luminosity of $818~\mathrm{pb^{-1}}$ and approximately $5.4$ million events, and they found $|V_{cd}|=0.234\pm 0.007\pm 0.002\pm 0.025$~\cite{CLEO:2009svp}. In the past 10 years, the BESIII Collaboration has done a lot of relevant studies on the CKM matrix element $|V_{cd}|$ under different decay channels. In 2015, in an an analysis of a $2.92~\mathrm{fb^{-1}}$ data sample taken at $3.773~\mathrm{GeV}$ with the BESIII detector operated at the BEPCII, the BESIII Collaboration measured the absolute decay branching fractions to be $\mathcal{B}(D^0\to K^-e^+\nu_{e})$ and $\mathcal{B}(D^0\to \pi^- e^+ \nu_{e})$. At the same time, they extracted the value of the CKM matrix element of $|V_{cd}|=0.2155\pm 0.0027\pm 0.0014\pm0.0094$~\cite{BESIII:2015tql}. As recently as 2020, the BESIII Collaboration measured for the first time the absolute branching fraction of the $D^+\to \eta \mu^+\nu_{\mu}$ decay and obtained $|V_{cd}|=0.242\pm0.022_{\mathrm{stat}}\pm0.006_{\mathrm{syst}}\pm0.033_{\mathrm{theory}}$~\cite{Ablikim:2020hsc}. Compared to semileptonic decay, the leptonic decay is a cleaner and simpler channel to extract the CKM matrix element $|V_{cd}|$. In 2013, the BESIII Collaboration found the value for the CKM matrix element $|V_{cd}|=0.2210\pm0.0058\pm0.0047$ through using the measurement of the branching fraction $\mathcal{B}(D^+\to\mu^+\nu_{\mu})$ together with a theoretical prediction for $f_{D^+}$~\cite{BESIII:2013iro}. In 2019, the BESIII Collaboration reported the first observation of $D^+\to\tau^+\nu_{\tau}$ with a significance of $5.1\sigma$, and obtained $|V_{cd}|=0.237\pm 0.024_{\mathrm{stat}}\pm 0.012_{\mathrm{syst}}\pm 0.001_{\mathrm{ex-syst}}$ by measuring $\mathcal{B}(D^+\to \tau^+\nu_{\tau})=(1.20\pm 0.24_{\mathrm{stat}\pm 0.12_{\mathrm{syst}}})$~\cite{BESIII:2019vhn} and using external inputs. The accurate extraction of the CKM matrix elements from experimental group is difficult because of the uncertainty in the measurement of the decay constant.

On the other hand, the charmed meson leptonic decay are the golden channel searching for the decay constant. Since 2005, numerous theoretical and experimental works have been made for improving the study of the decay constants. Experimentally, new data on the charmed mesons decay constant $f_{D^+}$ has been reported. Early in 2005, the CLEO Collaboration adopted 281 $pb^{-1}$ of data taken on the $\psi(3770)$ resonance with the CLEO-c detector to extract the value for the $D^+$-meson decay constant $f_{D^+}=(222.6\pm 16.7_{-3.4} ^{+2.8})~\mathrm{MeV}$ by measuring the branching fraction for $\mathcal{B}(D^+ \rightarrow  \mu^+ \nu_{\mu})$~\cite{CLEO:2005jsh}. Then the CLEO Collaboration used 818 $pb^{-1}$ of data taken on the $\psi(3770)$ resonance with the CLEO-c detector at the CESR collider to measure the branching fraction for the purely leptonic decay of the $D^+$-meson $\mathcal{B}(D^+ \rightarrow  \mu^+ \nu_{\mu}) = (3.82 \pm 0.32 \pm 0.09) \times 10^{-4}$, and utilized the determination to derive the higher precision value for the pseudoscalar decay constant $f_{D^+} = (205.8 \pm 8.5 \pm 2.5) ~\mathrm{MeV}$ in 2008~\cite{CLEO:2008ffk}. Meanwhile the BESIII Collaboration measured the value for the branching  fraction $\mathcal{B}(D^+ \rightarrow  \mu^+ \nu_{\mu}) = [3.71 \pm 0.19(\mathrm{stat}) \pm 0.06(\mathrm{sys})]\times 10^{-4}$,  with the CKM matrix element $|V_{cd}|$ determined from a global SM fit, which predicted the weak decay constant $f_{D^+}=(203.2\pm 5.3 \pm 1.8)~\mathrm{MeV}$  in 2013~\cite{BESIII:2013iro}. The Particle Data Group showed the average of the decay constant and the result was $f_{D^+}=212.0(7)~\mathrm{MeV}$ in 2024~\cite{ParticleDataGroup:2024cfk}. And Heavy Flavour Averaging Group (HFLAV) presented the average that was $f_D = (205.1\pm 4.4)~\mathrm{MeV}$ in 2022~\cite{HFLAV:2022esi}. The Flavour Lattice Averaging Group (FLAG) showed the mean value of $D$-meson which was $f_D=209.0(2.4)~\mathrm{MeV}$ in 2019~\cite{FlavourLatticeAveragingGroup:2019iem}. It can be seen that there are obvious differences in the measurements between different experimental groups, which also indicates that theoretical prediction is needed to do some evidence to obtain more accurate values.

Relative to the average presented by the average groups, it is difficult to reproduce the value consistently in theoretical calculations, such as the QCD sum rules (SR)~\cite{Wang:2015mxa,Dhiman:2017urn}, Lattice QCD~\cite{Collins:2024wab,Ke:2023qzc,Chen:2020qma,Bazavov:2017lyh,Boyle:2017jwu,Carrasco:2014poa, FermilabLattice:2014tsy, Dimopoulos:2013qfa, FermilabLattice:2013quj, FermilabLattice:2012lxl, FermilabLattice:2011njy}. In Ref.~\cite{Bazavov:2017lyh}, the calculation makes use of the highly-improved staggered-quark (HISQ) action for the sea and valence quark, in which the prediction for $B$ and $D$ mesons such as $f_{D^+}=212.7(0.6)~\mathrm{MeV}$, $f_{D_s}=249.9(0.4)~\mathrm{MeV}$, $f_{D_s}/f_{D^+}=1.1748(16)$, where the errors include statistical and all systematic uncertainties. In Ref.~\cite{Collins:2024wab}, the LQCD gives results obtained from numerous $N_f=2+1$ CLS ensembles which span six lattice spacings in the range $0.039~\mathrm{fm}\leqslant a\leqslant 0.1~\mathrm{fm}$ and lie on three trajectories in the quark mass plane, in which the decay constant for the $D$-meson read $f_D=208.4(0.67)_{\mathrm{stat}}(0.75)_{\mathrm{sys}}(1.11)_{\mathrm{scale}}[1.5]~\mathrm{MeV}$. Here we focus on one such method that is useful for solving nonperturbative problems of hadron physics: QCDSR. The discrepancies between the theoretical values and experimental data maybe signal some new physics beyond the SM. 

The QCD sum rules approach suggests to use the non-vanishing vacuum condensates to represent the non-perturbative effects~\cite{Shifman:1978bx}. The QCD background field approach provides a simple physical picture for those vacuum condensates from the viewpoint of field theory~\cite{Hubschmid:1982pa}. It assumes that the quark and gluon fields are composed of background fields describing the non-perturbative effects, while the quantum fluctuations represent the calculable perturbative effects. As a combination, the QCD sum rules within the framework of the background field theory (BFTSR) approach provides a clean physical picture for separating the perturbative and non-perturbative properties of the QCD theory and provide a systematic way to derive the QCD sum rules for hadron phenomenology which greatly simplifies the calculation due to its capability of adopting different gauges for quantum fluctuations and background fields. 
In order to give  different references for theoretial calculations, we use QCD sum rules to obtain the decay constant $f_{D^+}$ sum rules in this paper and further obtain branching fraction $\mathcal{B}(D^+\to\ell^+\nu_{\ell})$.
In which, the sum rule of $f_{D^+}$ can reduce the system uncertainties caused by the truncation of the high-dimensional condensates as well as the simple parametrization of quark-hadron duality for continuum states, and this improves the prediction accuracy of QCD sum rules.  

\begin{figure}[t]
\begin{center}
\includegraphics[width=0.4\textwidth]{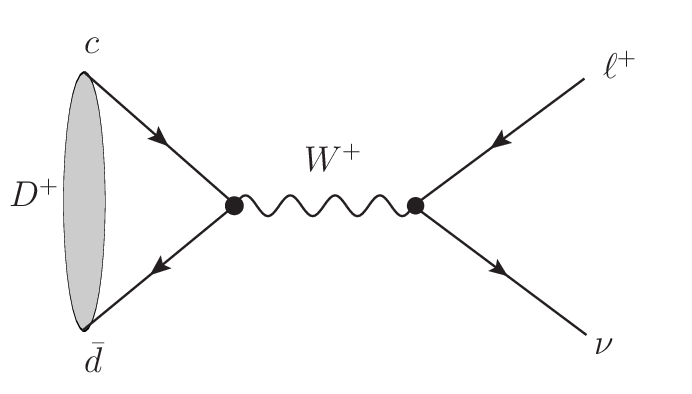}
\end{center}
\caption{The diagram for the leptonic decay $D^+ \to \ell^+ \nu $ through $c$ and $\bar{d}$ quarks annihilating into a pair of charged and neutral leptons via a virtual $W^+$-boson.}
\label{Fig:feynman}
\end{figure}

\textit{Theoretical Framework.--}The decay $D^+ \to \ell^+ \nu$ proceeds by the $c$ and $\bar{d}$ quarks annihilating into a virtual $W^+$-boson, the Feynman diagram for $D^+ \to \ell^+ \nu$ is shown in Fig.~\ref{Fig:feynman}. In the SM, the branching fraction for leptonic $D^+$-meson decay $D^+ \to \ell^+ \nu$ with $\ell = ( e,\mu, \tau)$ :
\begin{align}
&\mathcal{B}(D^+ \to \ell^+ \nu_\ell)
\nonumber\\
&\qquad= \frac{G^2_F}{8\pi} f^2_{D^+} m^2_\ell m_{D^+} \bigg( 1-\frac{m^2_\ell}{m^2_{D^+}} \bigg)^2 |V_{cd}|^2 \tau_{D^+}
\end{align}
where $G_F$ is Fermi coupling constant, $\tau_{D^+}$ is the lifetime of the $D^+$-meson, $m_{D^+}$ and $m_\ell$ are the mass of the $D^+$-meson and the lepton $(\ell = e, \mu, \tau)$, respectively. $|V_{cd}|$ is the magnitude of the $c \to d$ CKM matrix element. Given that all other variables are known with a high degree of precision, it is possible to obtain the branching fraction required by determining the $\Gamma(D^+ \to \ell^+\nu_\ell)$ directly and uniting the measured $f_{D^+}$.

In order to calculate the $D^+$-meson decay constant from the QCD sum rule approach, one should start from its basic definition with matrix element which have the following form
\begin{align}
\langle 0|\bar{c}(0)\gamma_\mu \gamma_5 d(0)|D^+(q) \rangle= i q_\mu f_{D^+}.
\label{eq:fDplus}
\end{align}
The above formula can also be get from the definition of twist-2 distribution amplitude $\xi$-moments of the lowest Fork state $D^+$-meson, {\it i.e.} $\langle 0| \bar{c}(0)\slashed{z} \gamma_5 (iz\cdot \tensor{D})^n d(0)|D^+(q)\rangle =i (z\cdot q)^{n+1} f_{D^+} \langle \xi^n\rangle_{D^+}$. When taking the order $n$ equals to zero, one can get the expression of Eq.~\eqref{eq:fDplus}. Further, we construct suitable two-point correlation function 
\begin{align}\label{correlation_function}
\Pi_{D^+}(z,q) &= i \int d^4xe^{iq\cdot x} \langle 0|T \left\{J(x)J^\dagger(0)\right\}|0 \rangle \nonumber \\
&= (z\cdot q)^{2} I_{D^+}(q^2),
\end{align}
with the interpolating currents
\begin{align}
\begin{split}
&J(x) = \bar{c}(x) \slashed{z} \gamma_5 d(x),\\
&J^\dagger(0)=\bar{d}(0)\slashed{z} \gamma_5 c(0).
\end{split}
\end{align}
According to the standard procedures of QCD sum rules, we firstly use the operator expansion (OPE) for the correlation function in the deep Euclidean region.
The calculation is carried out in the framework of BFT. 
By decomposing quark and gluon fields into classical background fields describing nonperturbative effects and quantum fields describing perturbative effects, BFT can provide clear physical images for the separation of long- and short-range dynamics in OPE.  
With the basic assumption of BFT and the corresponding Feynman rules, Eq.~(\ref{correlation_function}) can be rewritten as
\begin{align}\label{BFT}
\Pi_{D^+}(z,q) &= i \int d^4x e^{iq\cdot x}
\nonumber\\
&\times \bigg\{-\mathrm{Tr} \langle 0|S_F^c(0,x)\slashed{z}\gamma_5 S_F^q(x,0)\slashed{z}\gamma_5|0 \rangle\nonumber\\
&+\mathrm{Tr}\langle 0|\bar{c}(x)c(0)\slashed{z}\gamma_5 S_F^q(x,0) \slashed{z} \gamma_5 |0 \rangle\nonumber\\
&+\mathrm{Tr}\langle 0|S_F^c(0,x)\slashed{z}\gamma_5\bar{q}(0)q(x)\slashed{z}\gamma_5 |0 \rangle\bigg\}\nonumber\\
&+\cdot\cdot\cdot,
\end{align}
where ``Tr'' indicates trace for the $\gamma$-matrix and color matrix, $S_F^c(0,x)$ and $S_F^q(x,0)$ are quark propagators in the background field, respectively.
The tedious expressions of the propagators with terms leading to dimension-six condensates in the sum rules can be found in our previous work~\cite{Zhang:2017rwz}.
The first term in Eq.~(\ref{BFT}) gives perturbative contribution, involving the dimension-four gluon condensate and the dimension-six gluon condensate, respectively.
The other terms involve the dimension-three quark condensate $\langle \bar{q}q\rangle $, the dimension-five quark-gluon mixing condensate $\langle g_s\bar{q}\sigma TGq\rangle$ and the dimension-six quark condensate $\langle g_s\bar{q}q \rangle^2$. For the infrared divergence presented in the first term, we adopt the $D$-dimensional regularization approach to deal with the infrared divergence, $D=4-2\epsilon~(\epsilon\to 0)$. Furthermore, we apply the Feynman parametrization formula to obtain the key integral equation within it. 
\begin{align}
I(m,a,b,c) &=\sum_{k=0}^m \frac{(-1)^km!}{k!(m-k)!} \int_0^1dxx^{m-k-ak\epsilon}(1-x)^{k+b}\nonumber\\
           &\times \left( 1-\frac{q^2}{-q^2+m_c^2}x \right)^{-c-\epsilon},
\end{align}
It can be simplified with the help of the hypergeometric function,i.e.
\begin{align}
F(\alpha,\beta,\gamma,Z) &= \frac{\Gamma(\gamma)}{\Gamma(\beta)\Gamma(\gamma-\beta)}\nonumber\\
                         &\times \int_0^1dxx^{\beta-1}(1-x)^{\gamma-\beta-1}(1-Zx)^{-\alpha}\nonumber\\
                         &= \sum_{l=0}^{\infty} \frac{(\alpha)_l(\beta)_l}{l!(\gamma)_l}Z^l,
\end{align}
where $|Z|<1$ and $(\lambda)_l=\Gamma(\lambda+l)/\Gamma(\lambda)$, we can obtain
\begin{align}
I(m,a,b,c) &= \sum_{k=0}^m \frac{(-1)^km!}{k!(m-k)!} \frac{\gamma(k+b+1)}{\Gamma(c+\epsilon)}\nonumber\\
           &\times \sum_{l=0}^\infty \frac{\Gamma(l+c+\epsilon)\Gamma(l+m-k-a+1-\epsilon)}{l!\Gamma(l+m-a+b+2-\epsilon)}\nonumber\\
           &\times \left( \frac{-q^2}{-q^2+m_c^2}\right)^l.
\end{align}
The infrared divergence appears in $\Gamma (l+m-k-a+1-\epsilon)$ at the lowest several l-terms. We adopt the $\overline{\rm MS}$-scheme to deal with the divergent terms, which shall be absorbed into the renormalized $D^+$-meson decay constant $f_{D^+}$.
In addition, the correlation function Eq.~(\ref{correlation_function}) is obtained by inserting a completed set of intermediate hadronic states in the physical region.
The OPE of the correlation function Eq.~(\ref{correlation_function}) and its hadron expansion in deep Euclidean region can be matched by the dispersion relation. By further applying the Borel transformation for both sides, the QCD sum rule formula for the $D^+$-meson decay constant $f_{D^+}$ can be written as

\begin{widetext}
\begin{align}
f_{D^+}^2(M^2, s_0^{D^+}) &= M^2 e^{\frac{m_{D^+}^2}{M^2}}\bigg\{ \frac{1}{\pi} \frac{1}{M^2} \int_{t_{min}}^{s_0^{D^+}} ds e^{- \frac{s}{M^2}} \frac{1}{8\pi} \bigg\{ \left[ 2 \frac{m_c^2}{s} \left( 1- \frac{m_c^2}{s}\right) +1\right] \left( 1- \frac{2m_c^2}{s}\right)+1 \bigg\} -e^{-\frac{m_c^2}{M^2}} \frac{m_q \langle \bar{q}q\rangle}{M^4}
\nonumber
\\
&+\frac{\langle \alpha_s G^2 \rangle}{M^4} \frac{1}{12\pi} \left[ \mathcal{H}(0,0,0)-\frac{m_c^2}{M^2} \mathcal{H}(0,1,-2) \right]-e^{-\frac{m_c^2}{M^2}} \frac{m_q \langle g_s\bar{q}\sigma TGq\rangle}{M^6} \left[\frac{1}{18}+\frac{2m_c^2}{9M^2}\right]\!-\!e^{-\frac{m_c^2}{M^2}}\frac{\langle g_s\bar{q}q\rangle^2}{M^6}
\nonumber
\\
&\times\frac{2}{81} \,+\, \frac{\langle g_s^3fG^3\rangle}{M^6}\,e^{-\frac{m_c^2}{M^2}}\,\frac{1}{\pi^2}\,\bigg\{\,-\,\frac{17}{96}\mathcal{F}_1(0,5,3,2,\infty)
\,-\, \frac{1}{96}\mathcal{F}_2(0,4,3,1,\infty) \,+\, \frac{1}{144}\,\mathcal{F}_2(0,3,3,1,\infty)
 \nonumber
 \\
&+\frac{1}{288} \, \left[ 204\,+ \left( 100n\,-154\,+51\frac{m_c^2}{M^2}\right)\right] \,\left[\mathrm{ln}\frac{M^2}{\mu^2}+\psi(3)\right]+\frac{1}{288}\left( 17\frac{m_c^2}{M^2}-4n\right)\bigg\} \,+\frac{\langle g_s^3fG^3\rangle}{M^6}\frac{1}{\pi^2}
\nonumber
\\
&\times\bigg\{\frac{1}{288}\,\left[\,4\mathcal{H}(0,0,0)\,-\,3\mathcal{H}(0,0,-1) \,-\, 51\mathcal{H}(0,1,-2)\right] \,+\, \frac{1}{288}\frac{m_c^2}{M^2} [\, -\,2\mathcal{H}(0,0,-2)\,+\, 4\mathcal{H}(0,0,-1)
\nonumber
\\
&-2\mathcal{H}(-1,1,-2)-3\mathcal{H}(0,1,-3)]+\frac{1}{240}\frac{m_c^4}{M^4}\mathcal{H}(0,1,-4)\bigg\} \bigg\}
\label{fD+_function}
\end{align}
where $s_0^{D^+}$ is the continuous threshold parameter. For convenience, we present the expressions for every term in the decay constant (\ref{fD+_function}) in following
\begin{align}
&\mathcal{F}_1(n,a,b,l_{min},l_{max})=\sum_{k=0}^{n}\frac{(-1)^kn!\Gamma(k+a)}{k!(n-k)!}
                                     \sum_{l=l_{\mathrm{min}}}^{l_{\mathrm{max}}}\frac{\Gamma(l+b)\Gamma(n\!-\!1\!-\!k\!+\!l)}{\Gamma(n\!-\!1\!+\!l\!+\!a)}
                                     \sum_{i=0}^{l}\frac{1}{i!(l\!-\!i)!(l\!-\!1\!-\!i\!+\!b)!}\left(-\frac{m_c^2}{M^2}\right)^{l-i}
                                     \\
&\mathcal{F}_2(n,a,b,l_{min},l_{max})=\sum_{k=0}^{n}\frac{(-1)^kn!\Gamma(k+a)}{k!(n-k)!}
                                     \sum_{l=l_{\mathrm{min}}}^{l_{\mathrm{max}}}\frac{\Gamma(l+b)\Gamma(n-k+l)}{\Gamma(n\!+\!l\!+\!a)}
                                     \sum_{i=0}^{l}\frac{1}{i!(l-i)!(l\!-\!1\!-\!i\!+\!b)!}\left(-\frac{m_c^2}{M^2}\right)^{l-i}
                                     \\
&\mathcal{H}(n,a,b)=\int_0^1dx(2x-1)^nx^a(1-x)^be^{-\frac{m_c^2}{M^2(1-x)}}
\end{align}
\end{widetext}
More detailed explanation of the above formula can be found in our previous work ~\cite{Zhang:2017rwz}.
\\

\textit{Numerical Analysis.--}The numerical calculation is performed using the following parameters. According to the Particle Data Group (PDG) ~\cite{ParticleDataGroup:2024cfk},  we take the charm-quark mass $\bar{m}_c(\bar{m}_c)=1.28\pm0.03~\mathrm{GeV}$, the d-quark mass $\bar{m}_d(2~\mathrm{GeV})=4.7_{-0.4}^{+0.5}~\mathrm{MeV}$, the $D^+$-meson mass $m_{D^+}=1869.59\pm0.09~\mathrm{MeV}$. Furthermore, it is essential to ascertain the values of the non-perturbative vacuum condensates up to dimension-six, which include the double-quark condensates $\langle \bar{q}q\rangle$ and $\langle g_s\bar{q}q\rangle^2$, the quark-gluon condensate $\langle g_s\bar{q}\sigma TGq\rangle$, the double-gluon condensate $\alpha_{s} G^2$ and the triple-gluon condensate $\langle g_s^3fG^3\rangle$. We take their values from Refs.~\cite{Zhong:2014jla,Zhong:2021epq,Colangelo:2000dp}
\begin{align}
\langle \bar{q}q \rangle\mathrm{(2~GeV)}&=(-2.417_{-0.114}^{+0.227})\times10^{-2}~\mathrm{GeV^3},\nonumber\\
{\langle g_{s}\bar{q}q \rangle}^2\mathrm{(2~GeV)}&=(2.082_{-0.697}^{+0.734})\times10^{-3}~\mathrm{GeV^6},\nonumber\\
\langle g_{s}\bar{q}\sigma TGq \rangle\mathrm{(2~GeV)}&=(-1.934_{-0.103}^{+0.188})\times10^{-2}~\mathrm{GeV^5},\nonumber\\
\langle \alpha_{s}G^2 \rangle&=0.037\pm0.011~\mathrm{GeV^4},\nonumber\\
\langle g_{s}^3fG^3 \rangle&=0.045~\mathrm{GeV^6}.
\end{align}
\begin{figure}[t]
\begin{center}
\includegraphics[width=0.415\textwidth]{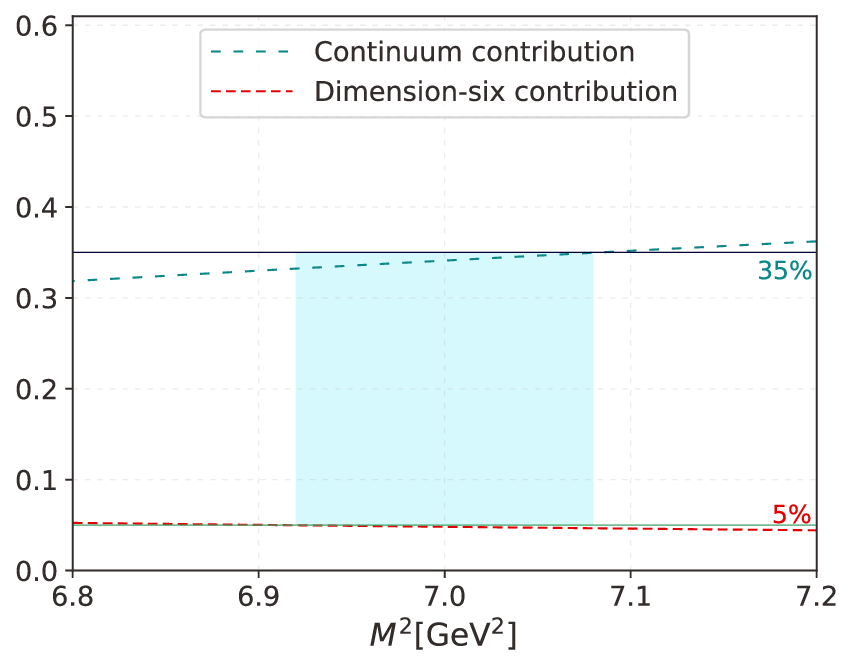}
\end{center}
\caption{Contributions from the continuum state and dimension-six condensates for the $D^+$-meson decay constant $f_{D^+}$ versus the Borel parameter $M^2$, where all input parameters are set to be their central values }
\label{Fig:borel-windows}
\end{figure}

\begin{figure}[t]
\begin{center}
\includegraphics[width=0.415\textwidth]{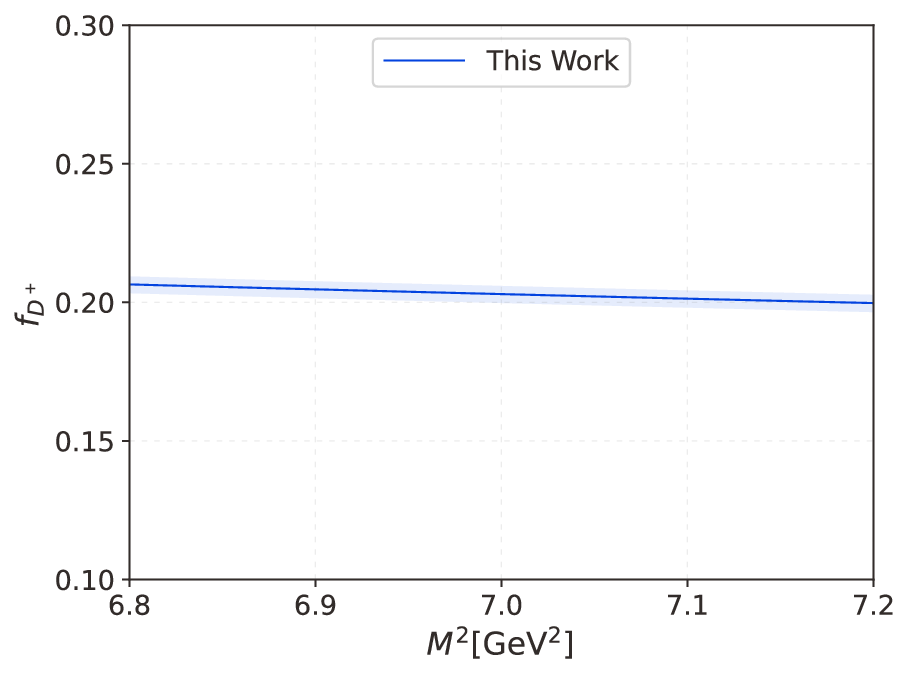}
\end{center}
\caption{The $D^+$-meson decay constant versus the Borel parameter $M^2$. The shaded band indicates the uncertainty.}
\label{Fig:FD}
\end{figure}

When doing the next step, the numerical calculation, every vacuum condensates and current quark masses need to be the required scale by applying the renormalization group equations (RGEs), as follow
\begin{align}
&m_d(\mu)=m_d(\mu_0)\left[\frac{\alpha(\mu)}{\alpha(\mu_0)}\right]^{\frac{4}{9}},\nonumber\\
&\bar{m}_c(\mu)=\bar{m}_c(\bar{m}_c)\left[ \frac{\alpha_s(\mu)}{\alpha_s(\bar{m}_c)}\right]^{-\frac{12}{25}},\nonumber\\
&\langle \bar{q}q \rangle(\mu)=\langle \bar{q}q \rangle(\mu_0)\left[ \frac{\alpha_s(\mu)}{\alpha_s(\mu_0)} \right]^{-\frac{4}{9}},\nonumber\\
&\langle g_s\bar{q}\sigma TGq \rangle(\mu)=\langle g_s\bar{q}\sigma TGq \rangle(\mu_0)\left[ \frac{\alpha_s(\mu)}{\alpha_s(\mu_0)} \right]^{\frac{2}{27}},\nonumber\\
&\langle g_s\bar{q}q \rangle^2(\mu)=\langle g_s\bar{q}q \rangle^2(\mu_0)\left[ \frac{\alpha_s(\mu)}{\alpha_s(\mu_0)} \right]^{-\frac{4}{9}},\nonumber\\
&\langle \alpha_s G^2 \rangle(\mu)=\langle \alpha_s G^2 \rangle(\mu_0),\nonumber\\
&\langle g_s^3fG^3 \rangle(\mu)=\langle g_s^3fG^3 \rangle(\mu_0).
\end{align}

In QCD sum rules analysis, the continuum threshold parameter $(s_0^{D^+})$ and the Borel parameter $M^2$ are both the important parameters. When calculating the decay constant $f_{D^+}$, one may set its continuum threshold to be close to the region around the square of the first excited state mass of $D^+$-meson, $D_0(2550)^0$ $\it{i.e}$, $s_0=5.4\mathrm{GeV^2}$. At the same time, to determine the Borel window, for the $D^+$ decay constant, we adopt the following criteria~\cite{Tian:2023vbh},  

\begin{itemize}
\item The continuum contribution is less than $35\%$;
\item The contribution of the six-dimensional condensates are no more than $5\%$;
\item The value of $f_{D^+}$ is stable in the Borel window;
\end{itemize}

Thus, the determined of the continuum and dimension-six condensates for $f_{D^+}$ are presented in Fig.~\ref{Fig:borel-windows}. In which, the shaded region stand for the Borel window. 
The continuum and the dimension-six condensates satisfy the above mentioned criteria, respectively, and the range of Borel window generated in them is reasonable. Moveover, it can be found in the Eq.~\eqref{fD+_function} that a large $M^2$ has a certain stabilizing effect on some the non-perturbative vacuum condensates up to dimension-six, $\it{e.g.}$ the quark-gluon condensate $\langle g_s \bar{q} \sigma TGq\rangle$, the double-quark condensates $\langle g_s\bar{q}q\rangle^2$ and the triple-gluon condensate $\langle g_s^3fG^3\rangle$.

\begin{figure}[t]
\begin{center}
\includegraphics[width=0.435\textwidth]{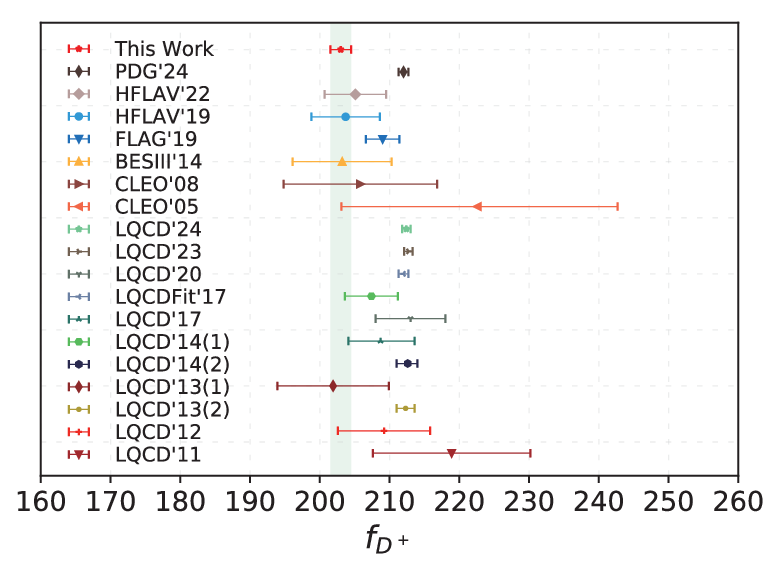}
\end{center}
\caption{We compared our own obtained predictions about $f_{D^+}$ with the results of other experimental and theoretical groups.}
\label{Fig:FDcomparison}
\end{figure}

\begin{table*}[t]
\footnotesize
\begin{center}
\caption{The branching fraction for $D^+ \rightarrow  \mu^+ \nu_{\mu}$, $D^+ \rightarrow  e^+ \nu_{e}$ and $D^+ \rightarrow  \tau^+ \nu_{\tau}$ channels. For comparison, the theoretical and experimental results are also presented.}
\label{Tab:BB}
\begin{tabular}{l l l l}
\hline
\hline
~~~~~~~~~~~~~~~~~~~~~~~~~~~~~~~~~~~~~~~~~~~~~~~&$\mathcal{B}(D^+ \rightarrow  e^+ \nu_{e} )$~~~~~~~~~~~~~&$\mathcal{B}(D^+ \rightarrow  \mu^+ \nu_{\mu} )(\times10^{-4})$~~~~~~~~~~~~~&$\mathcal{B}(D^+ \rightarrow  \tau^+ \nu_{\tau} )(\times10^{-3})$          \\
\hline
This Work                                      &$(8.260_{-0.118}^{+0.119})\times10^{-9}$          &$3.508_{-0.050}^{+0.051}$                                               &$0.935_{-0.013}^{+0.013}$
\\
LQCD'19~\cite{Fleischer:2019wlx}                 &$(9.16\pm0.22)\times10^{-9}$                      &$3.89\pm0.09$
&$1.04\pm0.03$
\\
BESIII'19~\cite{BESIII:2019vhn}                &-                                                 &-
&$1.2\pm0.24\pm0.12$
\\
BESIII'13~\cite{BESIII:2013iro}                &-                                                 &$3.71\pm0.19 \pm0.06$                                                                                                          &-
\\
CLEO'08~\cite{CLEO:2008ffk}                    &$<8.8\times10^{-6}$                               &$3.82\pm0.32\pm0.09$
&$<1.2$
\\
CLEO'06~\cite{CLEO:2006jxt}                    &-                                                 &-
&$<2.1$
\\
BESIII'05~\cite{BES:2004ufx}                   &-                                                 &$12.2_{-5.3}^{+11.1}\pm1.0$
&-
\\
CLEO'05~\cite{CLEO:2005jsh}                    &$<2.4\times10^{-5}$                               &$4.40\pm0.66_{-0.12}^{+0.09}$
&-
\\
CLEO'04~\cite{CLEO:2004pwu}                    &-                                                 &$3.5\pm1.4\pm0.6$
&-
\\
\hline
\hline
\end{tabular}
\end{center}
\end{table*}
In order to have a deep look at the decay constant $f_{D^+}$ changed with the Borel parameter $M^2$, we present the curves into Fig.~\ref{Fig:FD}. The shaded band stand for the uncertainties from the errors of all the mestioned input parameters including the quark mass, masons mass, each non-perturbative vacuum condensates. From the figure, we can see that the decay constant is flat in the allowable Borel windows which is consistent with the third criterion.
\begin{table}[t]
\footnotesize
\begin{center}
\caption{The results of $|V_{cd}|$ in different channels are compared, including the results of the experimental groups and the theoretical group. }
\label{Tab:BR}
\begin{tabular}{l l}
\hline
\hline
~~~~~~~~~~~~~~~~~~~~~~~~~~~~~~~~~~~~~~~~~~~~~~&$|V_{cd}|$                                            \\
\hline
This Work                                                 &$0.227_{-0.001}^{+0.002}$                               \\
PDG'24~\cite{ParticleDataGroup:2024cfk}                   &$0.221 \pm 0.004$
\\
LQCDFit'23~\cite{Ke:2023qzc}                                &$0.22486_{-0.00069}^{+0.00069}$                          \\
FLAG'21~\cite{FlavourLatticeAveragingGroupFLAG:2021npn}   &$0.2179 \pm 0.0057$                                       \\
BEIII'20~\cite{Ablikim:2020hsc}                           &$0.2422 \pm0.041 \pm 0.034$                                \\
FLAG'19~\cite{FlavourLatticeAveragingGroup:2019iem}       &$0.2116 \pm0.0050$                                          \\
BESIII'19~\cite{BESIII:2019vhn}                           &$0.237\pm 0.024\pm 0.12\pm0.001$
\\
BESIII'18(I)~\cite{BESIII:2018xre}                        &$0.217 \pm0.026 \pm0.004$                                    \\
BESIII'18(II)~\cite{BESIII:2018eom}                       &$0.2264 \pm0.0338 \pm0.0318$                                  \\
BESIII'17~\cite{BESIII:2017ylw}                           &$0.2243 \pm0.0058 \pm0.0026$                                   \\
BESIII'15~\cite{BESIII:2015tql}                           &$0.2278 \pm0.0034 \pm0.0023$                                    \\
BESIII'13~\cite{BESIII:2013iro}                           &$0.2210\pm 0.0058\pm 0.0047$
\\
CLEO'09~\cite{CLEO:2009svp}                               &$0.234 \pm0.007 \pm0.025$                                        \\
CKMfitter'04~\cite{Charles:2004jd}                        &$0.22487_{-0.00021}^{+0.00024}$                                    \\
\hline
\hline
\end{tabular}
\end{center}
\end{table}

By using the above three criteria and the chosen continuum threshold parameter we can get the numerical result of $f_{D^+}$, {\it i.e.}
\begin{align}
f_{D^+}=203.0\pm1.5~\mathrm{MeV},
\end{align}
which is also listed in Fig.~\ref{Fig:FDcomparison}. As a comparison, we also listed the experimental and theoretical predictions, e.g, PDG in 2024~\cite{ParticleDataGroup:2024cfk}, HFLAV in 2022~\cite{HFLAV:2022esi} and 2019~\cite{HFLAV:2019otj}, FLAG in 2019~\cite{FlavourLatticeAveragingGroup:2019iem}, BESIII in 2014~\cite{BESIII:2013iro}, CLEO Collaboration in 2008~\cite{CLEO:2008ffk} and 2005~\cite{CLEO:2005jsh}, LQCD~\cite{Collins:2024wab, Ke:2023qzc, Chen:2020qma, Bazavov:2017lyh, Boyle:2017jwu, Carrasco:2014poa, FermilabLattice:2014tsy, Dimopoulos:2013qfa, FermilabLattice:2013quj, FermilabLattice:2012lxl, FermilabLattice:2011njy}.  From the figures we can see that our prediction is in good agreement with the BESIII'14~\cite{BESIII:2013iro} prediction and the LQCD'13(1)~\cite{Dimopoulos:2013qfa} within errors. The uncertainties of the $f_{D^+}$  are caused by different input parameters. The differences and uncertainties arising from various groups' predictions, as well as our own predictions due to the application of different techniques, indicate the need for us to continue exploring the $D^+$-meson decay constant in pursuit of greater accuracy. It is expected that in the future we will obtain more accurate predictive values.

Furthermore, by determining $f_{D^+}=203.0\pm1.5~\mathrm{MeV}$ from the application of $f_{D^+}$ sum rules, we can further calculate the physical observations of leptonic decay $D^+ \to \ell^+ \nu$ with $\ell=(\mu,e,\tau)$. Thus, we can get the integrated values of leptonic $D^+$ decay widths in three channels:
\begin{align}
&\Gamma(D^+ \to e^+\nu_e) = (5.263_{-0.075}^{+0.076}) \times 10^{-21} ~\mathrm{GeV},\nonumber\\
&\Gamma(D^+ \to \mu^+\nu_{\mu}) = (2.236_{-0.032}^{+0.032}) \times 10^{-16} ~\mathrm{GeV},\nonumber\\
&\Gamma(D^+ \to \tau^+\nu_{\tau}) = (5.958_{-0.085}^{+0.086}) \times 10^{-16} ~\mathrm{GeV}.
\label{decaywidth}
\end{align}
Meanwhile, we can obtain the branching fraction results through the utilization of the lifetime of the initial state $D^+$-meson, ${\it i.e.}$ $\tau_{D^+} = 1.033 \pm 0.005~\mathrm{ps}$~\cite{ParticleDataGroup:2024cfk}. The comparison of prediction results are shown in the Table~\ref{Tab:BB}, which also includes comparison of experimental and theoretical results, {\it e.g.} the SM in 2019~\cite{Fleischer:2019wlx}, BESIII Collaboration in 2019~\cite{BESIII:2019vhn}, 2014~\cite{BESIII:2013iro}, 2005~\cite{BES:2004ufx} CLEO Collaboration in 2008~\cite{CLEO:2008ffk} and 2006~\cite{CLEO:2006jxt}, 2005~\cite{CLEO:2005jsh} and 2004~\cite{CLEO:2004pwu}. Through comparative observation, it is not difficult to find that our prediction results are roughly consistent with the experimental results. The uncertainties of $\mathcal{B}(D^+\to\ell^+\nu_\ell)$ caused by different input parameters such as $f_{D^+}$. The uncertainty in the system is still large, so we look forward to obtaining more accurate experimental results in the future.

In addition, a new result on $|V_{cd}|$ is obtained, we obtain the CKM matrix element $|V_{cd}|$ from the branching fractions $\mathcal{B}(D^+ \rightarrow  \mu^+ \nu_{\mu} )$ in the BESIII'13~\cite{BESIII:2013iro}. Our predictions are in Table~\ref{Tab:BR} and compared with theoretical groups of different experimental groups, where the errors are caused by all the mentioned error sources and the PDG errors for the branching fractions and the decay lifetimes. And our theoretical prediction aligns well with the experimental outcomes. Our theoretical prediction for $|V_{cd}|$ agree with that of the  BESIII'18(II)~\cite{BESIII:2018eom} within errors.
\\

\textit{Summary.--}In this paper, the $D^+$-meson decay constant $f_{D^+}$ sum rules have been obtained by using the QCD sum rules analysis expression based on the background field theory. In which, we improve the accuracy of the $f_{D^+}$ by considering the contribution of the six-dimensional condensates. The $f_{D^+}$ are obtained by constraining the contribution of continuum and dimension-six condensates, meanwhile, we obtain the behavior of the $f_{D^+}$ in the small $M^2$ region as shown in Fig.~\ref{Fig:FD}. In addition, based on the obtained $f_{D^+}$, we have also calculated the integrated values of leptonic decay widths \eqref{decaywidth} and the branching fraction in the Table~\ref{Tab:BB} of $D^+ \to \ell^+ \nu$ with $\ell=(\mu,e,\tau)$ lepton decays. Finally, we extracted the popular CKM matrix element $|V_{cd}|$ using the branching fractions $\mathcal{B}(D^+ \to \mu^+ \nu_\mu)$ in BESIII'13~\cite{BESIII:2013iro}, which is $|V_{cd}| = 0.227_{-0.001}^{+0.002}$.
\\

\textit{Acknowledgments.--}Tao Zhong and Hai-Bing Fu would like to thank the Institute of High Energy Physics of Chinese Academy of Sciences for their warm and kind hospitality. This work was supported in part by the National Natural Science Foundation of China under Grant No.12265009 and 12265010, the Project of Guizhou Provincial Department of Science and Technology under Grant No.ZK[2023]024.

\end{document}